\documentclass[twocolumn,showpacs,preprintnumbers,amsmath,superscriptaddress,amssymb,prl]{revtex4-1}
\usepackage{graphicx,latexsym,xcolor}
\usepackage{dcolumn}
\usepackage{bm}
\usepackage{booktabs}
\usepackage{float}
\usepackage{subfigure}

\begin{document}

\title{Topological phase transitions in an inverted InAs/GaSb quantum well driven by tilted magnetic fields}
\author{Hsiu-Chuan Hsu}
\affiliation{Department of Physics, National Taiwan University, Taipei 10617, Taiwan}
\author{Min-Jyun Jhang}
\affiliation{Department of Physics, National Taiwan University, Taipei 10617, Taiwan}
\author{Tsung-Wei Chen}
\affiliation{Department of Physics, National Sun Yat-sen University, Kaohsiung 804, Taiwan}
\author{Guang-Yu Guo}
\affiliation{Department of Physics, National Taiwan University, Taipei 10617, Taiwan}
\affiliation{Physics Division, National Center for Theoretical Sciences, Hsinchu 30013, Taiwan}

\date{\today}

\begin{abstract}
The helical edge states in a quantum spin Hall insulator are presumably protected by time-reversal symmetry.  
However, even in the presence of magnetic field which breaks time-reversal symmetry, the helical edge conduction can still exist, dubbed as pseudo quantum spin Hall effect. 
In this work, the effects of the magnetic fields on the pseudo quantum spin Hall effect and the phase transitions are studied. 
We show that an in-plane magnetic field drives a pseudo quantum spin Hall state to metallic state at a high field.
Moreover, at a fixed in-plane magnetic field, an increasing out-of-plane magnetic field leads to a reentrance of pseudo quantum spin Hall state in an inverted InAs/GaSb quantum well. 
The edge state probability distribution and Chern numbers are calculated to verify that the reentrant states are topologically nontrivial. 
The origin of the reentrant behavior is attributed to the nonmonotonic bending of Landau levels and the Landau level mixing caused by the orbital effect induced by the in-plane magnetic field. 
The robustness to disorder is demonstrated by the numerically calculated quantized conductance for disordered nanowires within Landauer-B\"uttiker formalism. 
\end{abstract}

\maketitle

\section{I. Introduction}
One of the important breakthroughs in condensed matter physics in recent years is the topological classification of the states of matter \cite{Schirber2016}. 
A bulk insulator can be classified by the topological properties of its band structure \cite{Hasan2010a, Qi2011}. 
One prominent distinction of a topologically nontrivial and trivial insulator is that the former possesses gapless states that appear at the interface with vacuum. 
The first discovered topological state of matter is quantum Hall (QH) insulator \cite{Klitzing1980a} in which the Hall conductance is quantized in units of $\frac{e^2}{h}$ as a consequence of 
the formation of Landau levels (LLs) under a magnetic field. 
The quantization integer has been revealed to be a topological invariant that describes the topological properties of the energy bands \cite{Thouless1982a} and have been called Chern number. 
At the interface between a QH insulator and vacuum, the edge states flow unidirectionally and thus are dissipationless. 

For a QH insulator, time-reversal (TR) symmetry is obviously broken in the presence of magnetic fields. 
Nonetheless, a few decades later, a TR symmetry preserved topological state has been discovered, the quantum spin Hall (QSH) insulator \cite{Kane2005a,Bernevig2006d,Konig2007a} 
originating from spin-orbit coupling and band inversion. 
In a QSH insulator, each spin sector possesses a quantized Hall conductance opposite to each other leading to a vanishing Hall conductance, which is the sum of that of both spin sectors.
As expected in any TR symmetric system, the Hall conductance must be zero.
Therefore, Chern number is not a useful topological characterization of a QSH insulator. 
Instead, it has been shown that the topological properties of a QSH insulator can be described by the $Z_2$ invariant \cite{Kane2005b} or the spin Chern number \cite{Sheng2006a}, 
which gives the equivalent topological description under TR symmetry despite of the different formulation \cite{Fukui2007a}. 

Along the sample boundary of a QSH insulator, there exhibits a pair of spin-polarized counterpropagating (helical) currents, for which the backscattering is expected to be prevented by TR symmetry. 
However, as pointed out in several theoretical works \cite{Tkachov2010,Yang2011,Scharf2012,Chen2012,Tkachov2012,Zhang2014, Scharf2015,Durnev2016, Hu2016}, the QSH effect is not destroyed immediately by the application of TR symmetry breaking terms.
A few theoretical studies have been devoted to understanding the robustness of the QSH effect under broken TR symmetry. 
In a Kane-Mele model with magnetization, the QSH effect and edge states survive till the bulk gap closes and reopens \cite{Yang2011}. 
For a QSH insulator in the presence of an out-of-plane magnetic field, 
the spin Chern number is shown to remain unchanged up to the magnetic field when the Landau levels cross \cite{Chen2012, Beugeling2012a,Zhang2014}. 
While for an inverted InAs/GaSb quantum well in the presence of an in-plane magnetic field, the edge states and the quantized conductance are shown to be robust till a strong magnetic field of $20$ T \cite{Hu2016}.
In a recent experiment, it was shown that the helical edge conduction survives up to an in-plane magnetic field of $12$ T \cite{Du2015}.
Several names have been coined to name this effect, for which the spin Chern numbers are integers in the presence of TR symmetry breaking, such as weak QSH effect \cite{Goldman2012}, TR-symmetry-broken QSH effect \cite{Yang2011}. 
Here, we follow \cite{Chen2011} and use {\it the pseudo QSH effect} to refer to this phenomenon. 


In this paper, we study the robustness of the pseudo QSH effect and the helical edge conduction in the presence of magnetic fields and disorder. 
In particular,  we investigate the magnetic field-driven topological phase transitions.
The widely-used Bernevig-Hughes-Zhang (BHZ) model is adapted to describe the QSH insulator with experimentally accessible parameters for an inverted InAs/GaSb quantum well.
By virtue of the electron and hole layer separation in an InAs/GaSb quantum well, an in-plane magnetic field gives rise to the orbital effect.
We show that an in-plane magnetic field can drive the inverted quantum well from pseudo QSH to metallic state. 
Moreover, as the out-of-plane magnetic field increases at a fixed in-plane magnetic field, at a certain Fermi energy regime, a reentrant pseudo QSH is uncovered which exists even in the presence of disorder. 
We attribute the reentrant pseudo QSH effect to the nonmonotonic behavior of Landau levels (LLs) due to weak electron-hole coupling and LL mixing by the in-plane magnetic field.
By tuning the Fermi energy to other regimes, the pseudo QSH to QH phase transition can also be observed. 
 
 
The rest of the paper is organized as follows. In Sec. II, the model and the formalism are given. 
In Sec. III, we present our results along with discussion, particularly the magnetic field-driven topological phase transitions and the effect of disorder and Zeeman field.
Spin Chern number, band structures and wave function distribution are presented for the characterization of the topological phases.
Finally, conclusions drawn from this work are given in Sec. IV.


\section{II. Model Hamiltonian}
 
To describe a QSH insulator, we apply the effective BHZ model, in the basis {$|E+\rangle, |H+\rangle, |E-\rangle, |H-\rangle$}, 
\begin{eqnarray}
H_o&=&
	\left(
\begin{array}{cc}
 h_+(k) &    0 \\
 0 &     h_-(k) 
\end{array}
\right),
\label{hamiltonian}
\end{eqnarray}
where $h_{\pm}(k)=-Dk^2 I+(M-Bk^2)\sigma_z\pm Ak_x\sigma_x-Ak_y\sigma_y$ with $\sigma_{x,y,z}$ the Pauli matrices and $I$ the $2\times 2$ identity matrix,  
$\pm$ denotes pseudo spin states and $E$ $(H)$ denotes electron (hole) states.
The parameters are chosen to describe an inverted InAs/GaSb quantum well: $B=-400$ meV nm$^2$, $D=-300$ meV nm$^2$, $A=23$ meV nm, $M=-8$ meV and the lattice constant ($a$) is $5$ nm \cite{Zhang2014}.
	\begin{figure}[htbp]
\begin{center}
\includegraphics[width=0.9\columnwidth]{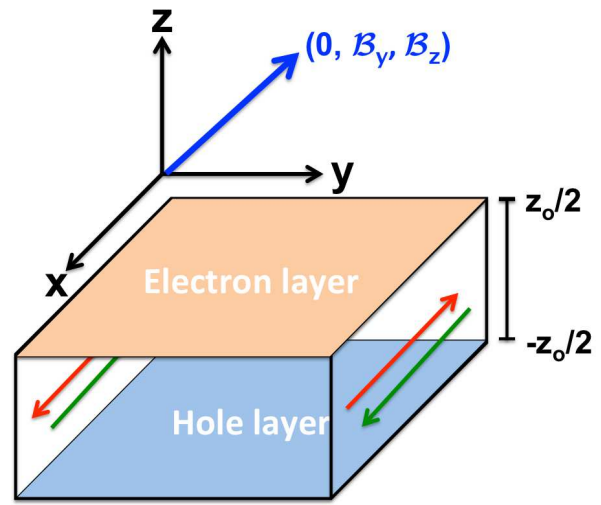}
\caption{ (Color online) Geometry of the setup. }
\label{setup}
\end{center}
\end{figure}
	
The quantum well lies on the xy-plane with its growth direction along the z-axis, as depicted in Fig. \ref{setup}.
A tilted magnetic field in the yz-plane $\vec{B}=(0,\mathcal{B}_y,\mathcal{B}_z)$ is applied.
In the presence of a magnetic field, the orbital motion of the carriers can be considered
in the Hamiltonian by Peierls substitution $\vec{k}\rightarrow \vec{k}-e\vec{A}$, where $\vec{A}$ is the 
vector potential and $e$ is the elementary charge \cite{dattabook, feynmanbook}. There is a freedom to choose the vector potential, and here we use the Landau gauge $\vec{A}=(-\mathcal{B}_zy+\mathcal{B}_yz,0,0)$. To model the spatial separation of the electron and hole for the bilayer system, we set the electron layer on $y=z_o/2$ plane, while the hole layer on the $y=-z_o/2$ plane, where $z_o$ is the thickness of the quantum well and taken to be one lattice constant \cite{Hu2016}. Thus, the vector potential for the electron(hole) layer is $\vec{A}=(-\mathcal{B}_zy+(-)\mathcal{B}_y\frac{z_o}{2},0,0)$.
When $\mathcal{B}_z$ is nonzero, translational invariance along $y$-direction is no longer preserved and $k_y$ is no longer a quantized number. 
We discretize the Hamiltonian along $y-$direction and rewrite the Hamiltonian as a one-dimensional (1d) tight binding model depending on $k_x$ \cite{Konig2008a}. 
The magnetic field is introduced as an phase factor $2\pi\frac{\mathcal{B}_zya}{\phi_o}$ felt by an electron when moving along the $x-$direction, where $\phi_o=h/e$ is the magnetic flux quantum.
To keep the periodicity along the y-direction, the magnetic flux is chosen to be a rational number and commensurate with the width $N_y$ \cite{Sheng2006b, Dutta2012}.

To determine the topological phases, we calculated the Hall conductance ($\sigma_H$). For an insulating phase, the Hall conductance is associated with the Chern number ($n$) as $\sigma_H=n\frac{e^2}{h}$, where $n$ characterizes the topological phase. 
The two pseudo spin components of the Hamiltonian are independent. 
Consequently, the energy bands and the Hall conductance $\sigma^{\pm}_H$ of each block Hamiltonian $h_{\pm}$ can be evaluated independently by the Kubo formula \cite{Bernevig2006d, Chen2011, Zhang2014}
\begin{equation}
\sigma_H^{\alpha}=\frac{-e^2}{\pi h}\sum_{{m\neq m'}}\int dk_x  \frac{Im\langle \Psi^{\alpha}_m| v^{\alpha}_x |\Psi^{\alpha}_{m'}\rangle \langle \Psi^{\alpha}_{m'}|v^{\alpha}_y |\Psi^{\alpha}_m\rangle}{(E^{\alpha}_m-E^{\alpha}_{m'})^2}, 
\end{equation}
where $\alpha=\pm$ denotes the pseodospin up and down block, $\Psi^{\alpha}_{m,m'}$ are the Bloch wave functions, $E^{\alpha}_{m(m')}$ are the occupied (unoccupied) energy bands and $v^{\alpha}_{x(y)}={-i}[\bold{r}_{x(y)},H]$ is the velocity operator with $\bold{r}$ the displacement operator. The Bloch wave function and the eigenenergies are obtained by direct diagonalization of the 1d tight binding Hamiltonian.
The total Hall conductance of the full Hamiltonian is given by $\sigma_H^{tot}=\sigma_H^++\sigma_H^-$, while the spin Hall conductance is given by $\sigma_s=\frac{h}{4\pi e}(\sigma_H^+-\sigma_H^-$) \cite{Beugeling2012a}. In insulating phases, the Hall conductances are proportional to the Chern number. Thus, the total Chern number is $n_{tot}=n_++n_-$, while the spin Chern number is given by $n_s=n_+-n_-$ \cite{Sheng2006a, Beugeling2012a}.
 In this study, we calculate the energy levels, the Hall conductance and the associated Chern number in the presence of a titled magnetic field.  
	
\section{III. Results and discussion}	
\subsection{A. The Pseudo QSH to Metallic Phase Transition}
\begin{figure}[htbp]
\begin{center}
\includegraphics[width=0.9\columnwidth, scale=0.6,angle=0]{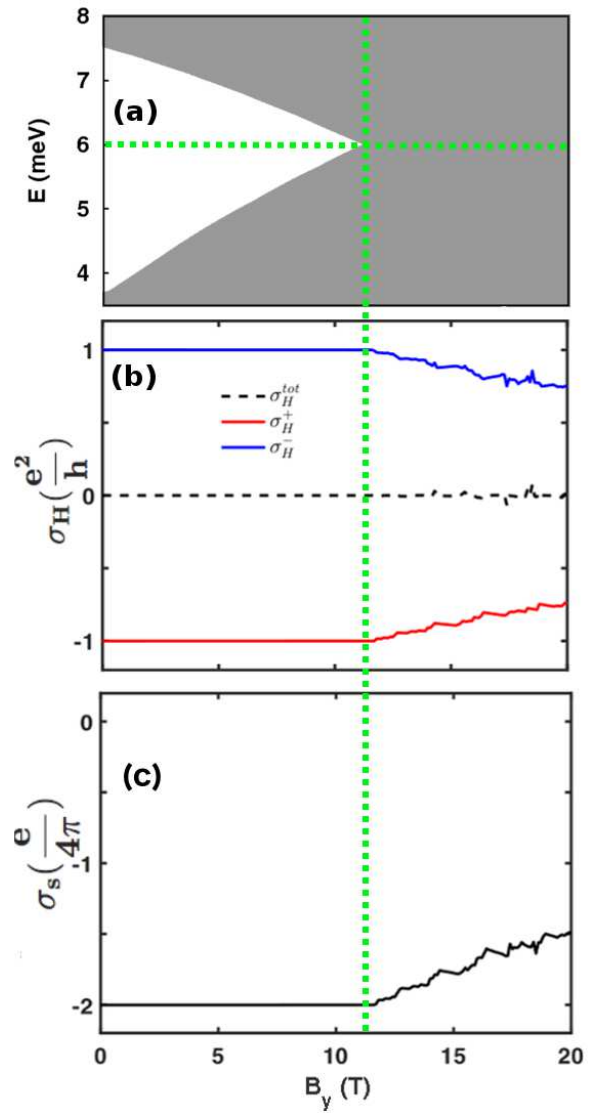}
\caption{(Color online) (a) Energy spectrum as a function of $\mathcal{B}_y$.  The horizontal dashed line denotes the Fermi energy $E_f$ for (b). (b) The corresponding Hall conductance for Fermi energy at $6$ meV. (c) The corresponding spin Hall conductance for Fermi energy at $6$ meV. The vertical dashed line denotes the transition point from pseudo QSH to metallic state.}
\label{inplanegap}
\end{center}
\end{figure}

First, we study the phase transition in the presence of an in-plane magnetic field, but without applying out-of-plane magnetic field. In the gauge chosen, after the Peierls substitution, the block Hamiltonian becomes 
\begin{eqnarray}
h_{\alpha}=
\left(
\begin{array}{cc}
M-(B+D)k_e^2 &  A(\alpha k_x+ik_y)  \\
 A(\alpha k_x-ik_y) &-M+ (D-B)k_h^2
%
\end{array}
\right),
\end{eqnarray}
 where $k_{e(h)}^2=(k_x+(-)\pi eB_yz_o /h)^2+k_y^2$.
 An explicit form of the Hamiltonian after the Peierls substitution can be found in the appendix.
We diagonalize the Hamiltonian and obtain the eigenenergies as a function of $\mathcal{B}_y$, as shown in Fig. \ref{inplanegap}(a). 
The energy spectrum remains gapped till $\mathcal{B}_{yc}=11.5$ T, above which the system becomes a metallic state. 
The Hall conductance $(\sigma^+_H,\sigma^-_H)$ at $E_f=6$ meV, as shown in Fig. \ref{inplanegap}(b), is constant and equals to $\frac{e^2}{h}(-1,1)$ till $\mathcal{B}_{yc}$, indicating a pseudo QSH phase. The integer ${-1,1}$ are Chern numbers that characterizes the topological property of the system. Our results agree well with Ref. \cite{Hu2016}, which shows that the helical edge states exist even in the presence of $\mathcal{B}_y$.  
In addition, the spin Hall conductance is shown in Fig. \ref{inplanegap}(c) and quantized for magnetic field up to $\mathcal{B}_{yc}$.
When $\mathcal{B}_y$ becomes larger than $\mathcal{B}_{yc}$, the system becomes metallic and the Hall conductance is no longer quantized. Similarly, as shown in Fig. \ref{inplanegap}(c), the spin Hall conductance gradually decreases. 

\subsection{B. The Reentrant Pseudo QSH Phase}
\begin{figure}[htbp]
\begin{center}
\includegraphics[width=0.9\columnwidth, scale=0.8]{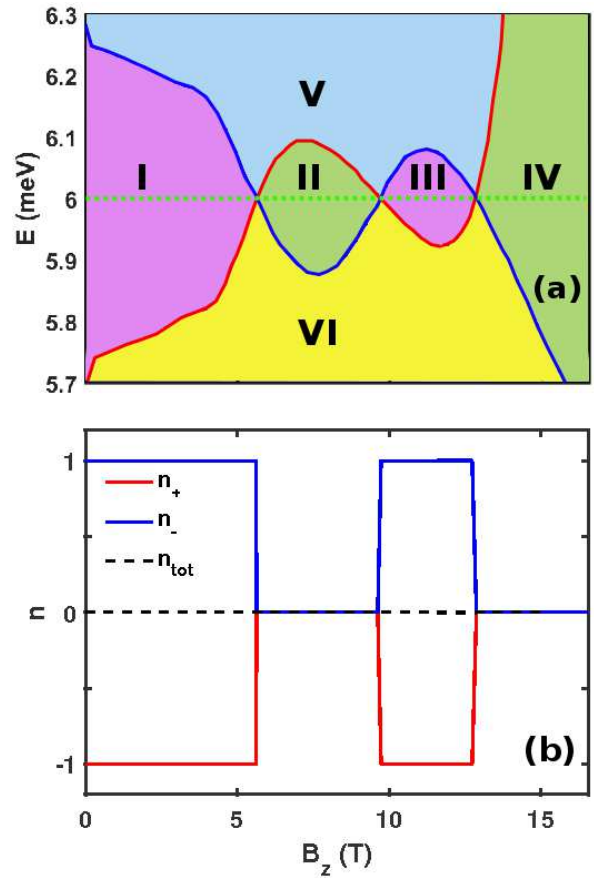}
\caption{(Color online) (a) Landau levels at $\mathcal{B}_y=10$ T. 
Regions V and VI are quantum Hall phases with Chern number 
$n_{tot} = -1$ and $n_{tot} = 1$, respectively, as explained in the text. 
The horizontal dashed line denotes the Fermi energy $E_f$ for (b).
(b) Chern number for each spin component for $E_f=6$ meV.}
\label{fig3}
\end{center}
\end{figure}

Next, we study the effect of tilted magnetic fields on the energy levels and the associated quantum phases. 
Fig. \ref{fig3} shows the energy spectrum and the associated Chern numbers for $\mathcal{B}_y=10$ T. 
In Fig. \ref{fig3}(a), we identify three quantum phase transitions by band crossing because two distinct topological phases cannot adiabatically connect to each other without closing bulk band gap. 
The topological nature of each phase is determined by the Chern number, as shown in Fig. \ref{fig3}(b). 
Thus, each region in Fig. \ref{fig3}(b) can be labeled by Chern numbers of each block Hamiltonian $(n_+,n_-)$: I (-1,1), II (0,0), III (-1,1), IV (0,0).
By fixing the Fermi energy near $6$ meV and increasing $\mathcal{B}_z$, the system undergoes transitions 
from a pseudo QSH state (I), a normal insulating state (II), the pseudo QSH state (III), 
and the trivial state(IV). The total Chern number ($n_{tot}$), which is the sum of both, remains zero. On the contrary, 
the spin Chern number, which is the difference between $n_{+}$ and $n_{-}$, goes from $-2$ to $0$,  
reenters $-2$, and $0$ again. 

\begin{figure*}[htbp]
\begin{center}
\includegraphics[width=0.95\textwidth, scale=1,angle=0]{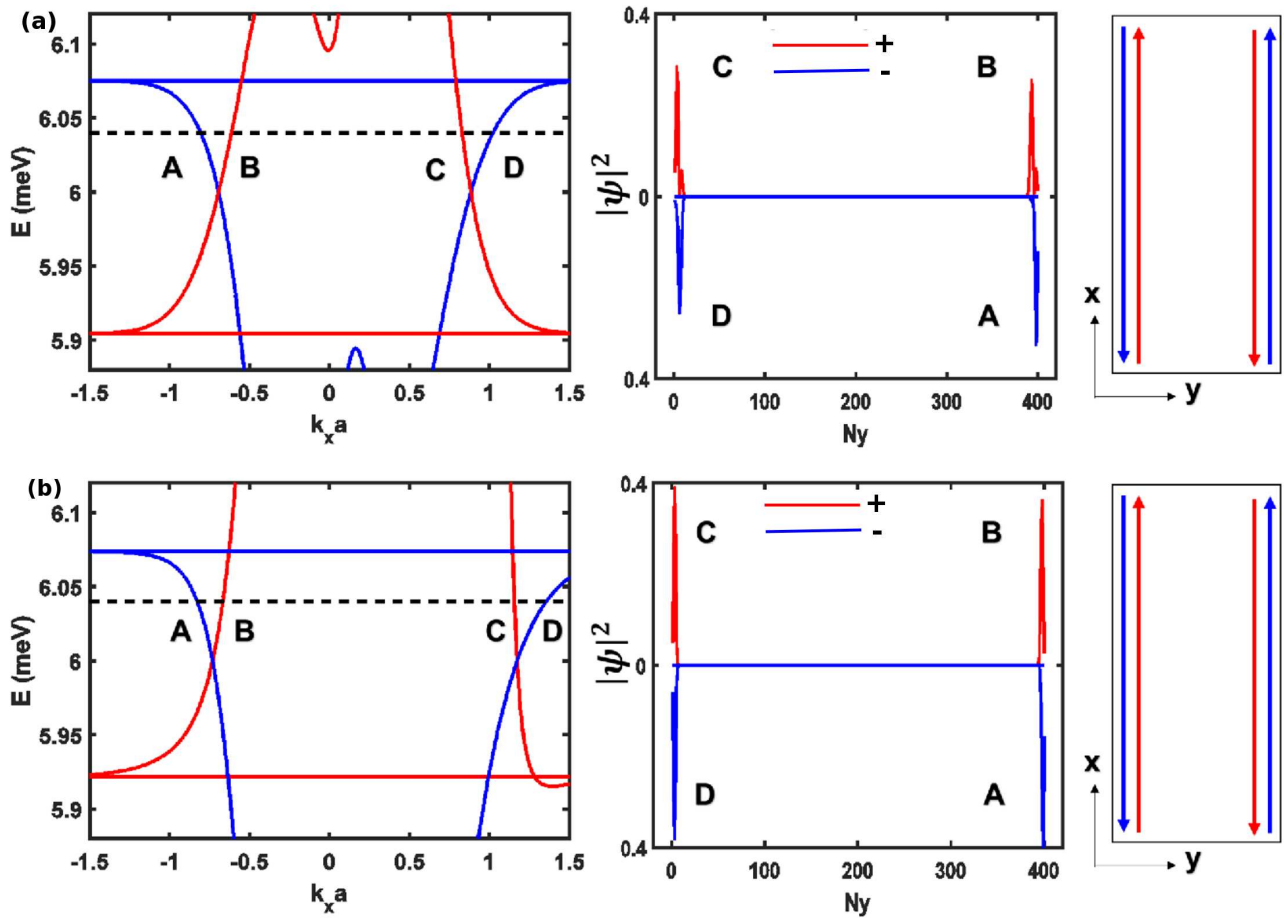}
\caption{(Color online) Energy band structure (left panel), edge state probability (middle panel) and current distribution 
(right panel) of a pseudo QSH nanoribbon at (a) $\mathcal{B}_z=5$ T in region I and (b) at $\mathcal{B}_z=11$ T 
in region III. }
\label{fig4}
\end{center}
\end{figure*}

In order to gain more insight into each pseudo QSH phase, we analyze the edge states in a ribbon geometry 
by imposing an open boundary condition along the $y-$direction.  
We study the energy dispersion and the edge state wave function in these two regions. Fig. \ref{fig4} gives the energy dispersion, probability distribution, and current flow in each region. 
Figures \ref{fig4}(a) and \ref{fig4}(b) correspond to regions I and III, respectively. 
The energy dispersions, as shown in the left panel in Fig. \ref{fig4}, show that there 
are gapless spectrum inside the gap of the Landau levels. 
The corresponding probability distribution of the gapless states $A,B, C,D$, as denoted in the energy dispersions, 
show that the gapless states are indeed edge states. Given that the electron velocity is 
given by $\frac{\partial E}{\hbar\partial k_x}$ and current by $I=-|e|v_x$, we obtain the edge current distribution 
in the right panel in Fig. \ref{fig4}. For the spin down states, state A has a negative velocity 
and lives on the right edge, while state D has a positive velocity and lives on the left edge. 
For the spin up states, state C has a negative velocity and lives on the left edge, 
while state B has a positive velocity and lives on the right edge.  In regions I and III, 
the direction of the spin current flow does not change. 
In the presence of magnetic field, the helical edge states are not destroyed immediately. 
Moreover, it is even possible to turn on and off the helical edge states by applying 
a tilted magnetic field without changing the direction of spin current.

To understand the theoretical reason of the reentrant behavior in the presence of tilted magnetic fields,  we turn to the analytical formulation of the LLs.
First, we review the solution of LLs in the presence of out-of-plane magnetic field only. 
The Hamiltonian can be simplified by defining ladder operators $a^{\dagger}$ and $a$, which raises and lowers 
the LL index by 1, respectively. The details of the solution are presented in the Appendix. 
The solution of the LLs is \cite{Konig2008a,Zhang2014}
\begin{eqnarray}
E^{n,s}_{\alpha}=n\omega_D+\frac{\alpha \omega_B}{2}+s\sqrt{(M+n\omega_B+\frac{\alpha\omega_D}{2})^2+\frac{2nA^2}{l_B^2}},  \notag\\
\label{LL}
\end{eqnarray}
where $\alpha=\pm$ denotes the pseodospin up and down block, $n$ is the LL index, $s=\pm$ denotes the eigenstate of each pseudospin block, $\omega_{B(D)}=-2B(D)/l_B^2$ and $l_B=\sqrt{\hbar/e\mathcal{B}_z}$. 
Since each block Hamiltonian is a $2\times 2$ matrix, every nonzero $n$th LL has two solutions with $s=\pm$ and each eigenstate is a 2-component vector.
For example, the eigenstate of the $h_+ (a^{\dagger},a)$ is $(f_{nl1}|n\rangle, f_{nl2}|n-1\rangle)$, where $f_{nl1(2)}$ is the normalization constant and $|n\rangle$ is the eigenstate of the number operator $a^{\dagger}a$. 
However, for the 0th LL, the eigenvector has only one nonzero component $(|0\rangle, 0)$ because $|0\rangle$ is already the lowest eigenstate of $a^{\dagger}a$. 
Thus, the solution of the 0th LL is  
\begin{eqnarray}
E_\alpha^0=\alpha M+\frac{\alpha\omega_B+\omega_D}{2}
\end{eqnarray}

\begin{figure}[htbp]
\begin{center}
\includegraphics[width=\columnwidth, scale=0.9,angle=0]{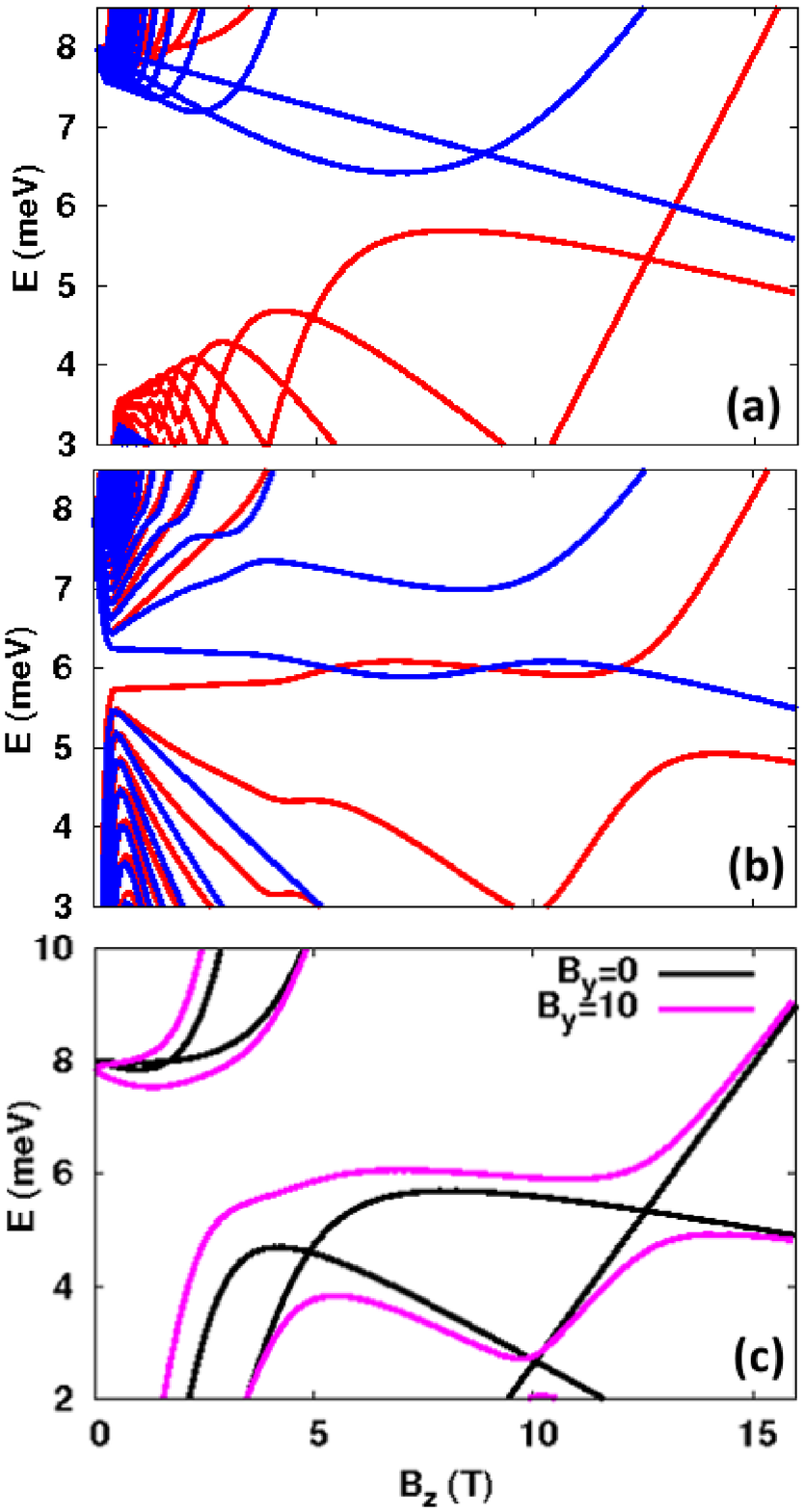}
\caption{(Color online) Energy levels for bulk using continuum model for $\mathcal{B}_y=0$ (a) and $\mathcal{B}_y=10 T$ (b). (c) Energy levels using the first three lowest LLs for spin up block Hamiltonian at $\mathcal{B}_y=10$ T(magneta) and at $\mathcal{B}_y=0$ T(black).  }
\label{fig5}
\end{center}
\end{figure}

Fig. \ref{fig5}(a) is the Landau level fan chart without in-plane magnetic field. There  is a pair of linear LL which are the 0th LL. 
The linear LL with positive slope is a pure $|E+\rangle$ state, while the one with negative slope is a pure $|H-\rangle$ state. 
The two 0th LLs cross at $\mathcal{B}_{zc}=13$ T beyond which the band sequence becomes normal. This implies that when $\mathcal{B}_z > \mathcal{B}_{zc}$, the pseudo QSH no longer exists, 
as verified with Chern number in the previous study \cite{Zhang2014}.



When both in-plane and out-of-plane magnetic fields are present, we treat the orbital effect from the in-plane magnetic field as an additional term and write the Hamiltonian with respect to the basis of the LLs at $\mathcal{B}_y=0$. 
After basis transformation, it is found that the in-plane magnetic field terms mix adjacent LLs and the Hamiltonian becomes tri-block diagonal.  We then obtain the energy levels in the presence of the nonzero $\mathcal{B}_y$ by direct diagonalization. The details of the method can be found in the Appendix.
Fig. \ref{fig5}(b) is the energy spectrum for $\mathcal{B}_y=10$T. There is no linear solutions as the zero modes in LLs due to the LL mixing by $\mathcal{B}_y$. 


The origin of reentrant behavior is two fold. First, before turning on $\mathcal{B}_y$, the LLs except zero modes are non monotonic. This is because of the spatial separation, the electron and hole layers are weakly coupled. 
The nonmonotonic relation with $\mathcal{B}_z$ is more obvious when the spin orbit coupling is vanishingly small. The eigenenergy can be written as 
$n\omega_D+\frac{\alpha \omega_B}{2}+s|(M+n\omega_B+\frac{\alpha\omega_D}{2})|$ in this limit. The last term in this expression is nonmonotonic because of the absolute value. 
This term leads to the band bending when the term in the absolute bracket equals to zero. In contrast, when spin-orbit coupling term is large, the LL change monotonically and cause no band bending. Thus, we do not expect such reentrance helical edge states to appear in HgTe quantum well which has stronger spin-orbit coupling.
Second, after turning on $\mathcal{B}_y$, the orbital effect mixes adjacent LLs and causes band splitting at the LL crossings. As shown in Fig. \ref{fig5}(c), the energy spectrum (magneta) of three LLs ($n=0,1,2$) for spin up block shows that the zero modes already open up a gap at LL crossings (black). 


\begin{figure}[htbp]
\begin{center}
\includegraphics[width=\columnwidth, scale=1,angle=0]{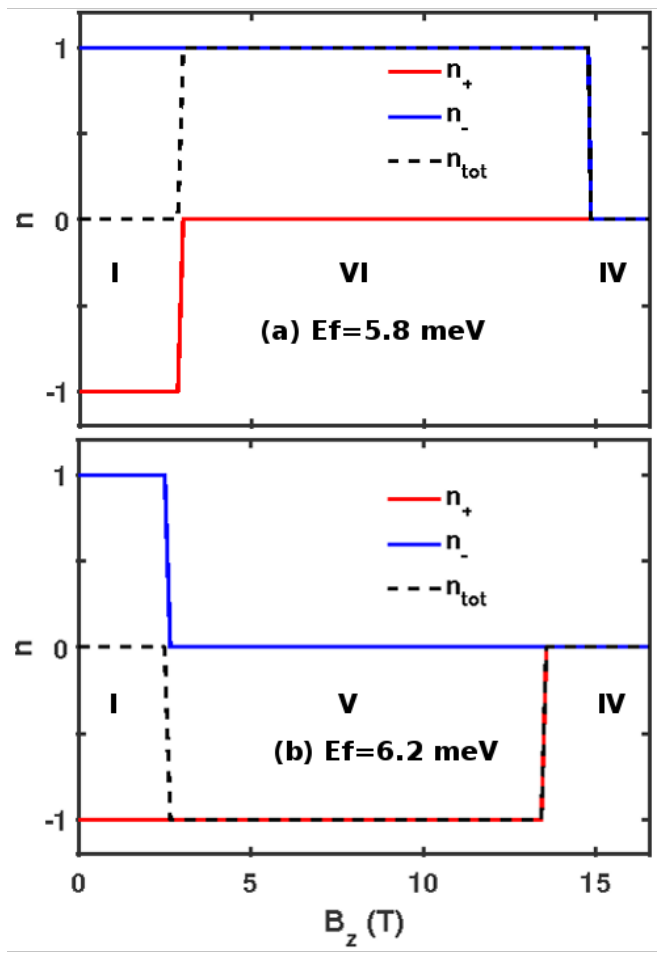}
\caption{(Color online) Chern number of each block Hamiltonian as a function of $\mathcal{B}_z$ at $\mathcal{B}_y=10$ T 
for $E_f=5.8$ meV (a) and $E_f=6.2$ meV (b).}
\label{qh}
\end{center}
\end{figure}
We have focused on the regime where the Fermi energy is inside the bulk gap. Nonetheless, 
once the Fermi energy is tuned away from $6$ meV, quantum Hall phases occur. In Fig. \ref{fig3}(a), 
two quantum Hall phases are shown:  V with $n_{tot}=-1$ and VI with $n_{tot}=1$. 
Figures \ref{qh}(a) and \ref{qh}(b) show the Chern number as a function of $\mathcal{B}_z$ at $E_f=5.8$ meV and $6.2$ meV, respectively. The system undergoes from pseudo QSH state to QH to trivial state as $\mathcal{B}_z$ increases. At the transitions involving QH, only one pseudospin species changes the phase, while the other remains the same, giving rise to a change of total Chern number by $1$. On the contrary, at the pseudo QSH to normal insulator transition, both pseudospin species undergoes the transition from topologically nontrivial to trivial insulator.

\subsection{C. Effects of Disorder and Zeeman Field}
\begin{figure}[htbp]
\begin{center}
\includegraphics[width=\columnwidth,angle=0,clip=false,scale=1]{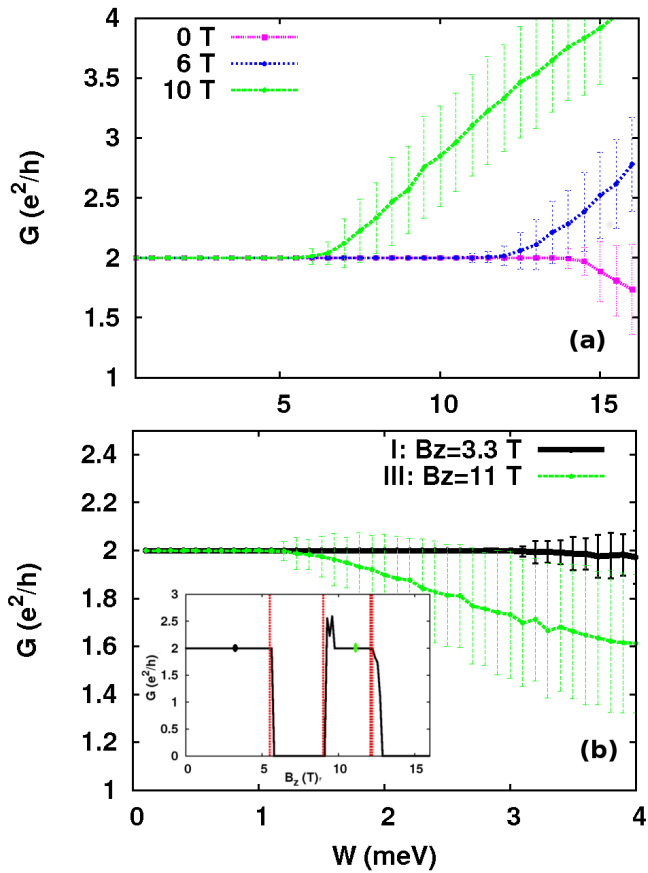}
\caption{(Color online) Two-terminal conductance for a nanoribbon of width 119a and length 200a at $E_f=6$ meV. 
The conductance is over 500 configurations. The standard deviation is plotted as error bars. 
(a) Conductance as a function of disorder strength for zero magnetic field (magenta), $\vec{B}=(0,6,0)$T (blue) 
and $\vec{B}=(0,10,0)$T (green). (b) Conductance for $\mathcal{B}_z=3.3$ T and $11$ T as $\mathcal{B}_y$ 
is fixed at $10 $T. The inset is conductance as a function of $\mathcal{B}_z$ in clean limit. 
The black (green) dot denotes $\mathcal{B}_z=3.3$ $(11)$ T. 
The red vertical lines denote where the energy levels cross at $\mathcal{B}_z=5.5, 9, 12.1$ T in Fig. 3(a).
}
\label{dis}
\end{center}
\end{figure}

Disorder is an unavoidable ingredient in realistic devices. 
In order to give a more complete picture of the quantum phase transition for experimentalist's interests, we study the robustness of the edge states in the presence of the on-site disorder. We rewrite the Hamiltonian Eq. (\ref{hamiltonian}) in two-dimensional tight-binding representation and introduce
the on-site disorder  as
 	\begin{eqnarray}
	H_{dis}&=&\left(
\begin{array}{cc}
h_{dis}  & 0    \\
0  &     h_{dis} 
\end{array}
\right)\notag\\
	h_{dis}&=&\sum_i
			\left(
\begin{array}{cc}
\epsilon_{ei}  &  0   \\
 0 &   \epsilon_{hi}  
\end{array}
\right),
	\end{eqnarray}
where $\epsilon_{e(h)i}=[-W/2,W/2]$ is the random onsite potential and $i$ is the site index.
The two-terminal conductance for a ribbon is calculated within Landauer-B\"uttiker formula with the recursive Green's function method. In the calculation, the width and the length of the nanoribbon is set to be 119a and 200a, respectively. The leads are modeled as normal metal without magnetic field penetrations.
We first study the robustness of the helical edge states in the presence of in-plane magnetic field only.  
Fig. \ref{dis}(a) shows the disorder averaged mean conductance and its standard deviation at $\mathcal{B}_y=0, 6$ and $10$ T. The critical disorder $W_c$ at which the conductance is no longer quantized decreases as $\mathcal{B}_y$ increases. This can be understood as the decrease of the energy gap as the in-plane magnetic field becomes stronger (Fig. \ref{inplanegap}(a)).
When $W<W_c$, the standard deviations are negligible, indicating that the helical edge states are perfect conducting.

When $\mathcal{B}_y$ is fixed at $10$ T, as shown in the inset of Fig. \ref{dis}(b), at $E_f=6$ meV, 
the conductance in clean limit starts at $2e^2/h$, drops to zero, then re-enters to $2e^2/h$ 
as $\mathcal{B}_z$ increases. The conductance suggests that in the ribbon structure, there are 
two conducting modes in regions I and III. The conductance as a function of disorder strength 
is calculated for  $\mathcal{B}_z=3.3$ T in region I and $\mathcal{B}_z=11$ T in region III.
The main figure of Fig. \ref{dis}(b) shows the mean conductance and the error bars 
over 500 disorder configurations. The quantized conductance at $\mathcal{B}_z=3.3$ T in region I 
and $\mathcal{B}_z=11$ T in region III remain up to a critical disorder strength
$W_c=3$ and $1.2$ meV, respectively. In this regime, where the conductance is quantized, 
the error bars are negligible, indicating the robustness of the conducting channels.  
For an approximately quantitative description of the robustness, we compare the critical 
disorder strength with the band gap. From Fig. \ref{fig3}(a), the band gap at these two magnetic fields 
are $0.37$ and $0.15$ meV, respectively. Consequently, the ratio of $W_c$ to the band gap is 
about $8$, indicating the robustness of the edge states in both regimes.



    An applied magnetic field would give rise to a vector potential which affects the orbital motion of the carriers and also a Zeeman field which may split or shift energy bands. While the former effect has been taken into account, the latter effect has so far been neglected. Nevertheless, recent experiments \cite{Du2015, Karalic2016} on the inverted InAs/GaSb quantum well bilayers have indicated that Zeeman effects are negligible and the helical edge states persist even in the presence of a strong magnetic field. In particular, Ref. \cite{Du2015} shows that the quantized plateau of conductance persists for $\mathcal{B}_y$ up to 12 T and $\mathcal{B}_z$ up to 8 T. There is no evidence that the Zeeman field opens a gap in the edge spectrum. In Ref. \cite{Karalic2016}, the authors demonstrate that the topological gaps persist as magnetic field strength increases up to $11$ T. Both facts suggest that the energy spectrum presented in Fig. 2(a) and the phase diagram displayed in Fig. 3(a) remain more or less valid even if the Zeeman effects would have been taken into account. Therefore, we conclude that in the considered InAs/GaSb quantum well bilayer, the Zeeman effects can be neglected. 


\section{IV. Conclusions}
In summary, we investigated the in-plane magnetic field effect on the the LL spectrum and topological phase transitions in an inverted InAs/GaSb quantum well. 
When driven only by an in-plane magnetic field, we found a pseudo QSH to metallic state transition at $\mathcal{B}_{yc} = 11.5$ T. 
More interestingly, at $\mathcal{B}_y$ fixed at $10$T, the quantum well exhibits a reentrance pseudo QSH effect as the out-of-plane magnetic field increases.
The origin of the reentrant behavior is the nonmonotonic bending of LLs due to the weak electron-hole coupling and the LL mixing.
Therefore, the reentrant pseudo QSH effect is not expected in HgTe quantum wells which has strong electron-hole coupling and no electron-hole layer separation. 
We also examined the robustness of the pseudo QSH effect against disorder for a nanoribbon structure. 
The critical disorder strength, at which the quantized conduction breaks down, is eight times of the bulk gap, indicating the robustness of the helical edge conduction in the presence of magnetic fields. 

\section{Acknowledgments}
This work is supported by the Academia Sinica, the Ministry of Science and Technology, 
the National Center for Theoretical Sciences of The R.O.C.

\appendix
\section{Appendix A: Peierls substitution}
In the presence of a vector potential, the momentum operator $-i\hbar \nabla$ is replaced by $-i\hbar \nabla - q\vec{A}$ in the continuum limit, where $q$ is the carrier charge \cite{dattabook, feynmanbook}. In the BHZ model, the Hamiltonian is based on electron picture. Thus, $q$ is taken to be electronic charge for both electron and hole layers. 

In the tight-binding framework, 
the carriers feel the vector potential in terms of the additional phase added to the hopping terms $t \rightarrow te^{i\theta}$, where $\theta=\frac{e}{\hbar}\int \vec{A}\cdot dl$ along the hopping path.
In the current model, for the hopping along the electron layer, the phase gained by the carrier is $\frac{e}{\hbar}\int B_y\frac{z_o}{2}\cdot dx$. While for the hopping along the hole layer, the phase gained by the carrier is $\frac{e}{\hbar}\int B_y\frac{-z_o}{2}\cdot dx$.
On the contrary,
for the off-diagonal terms in $h_{\pm}$ in Eq. (\ref{hamiltonian}) which correspond to the electron-hole coupling, the phase is zero because the origin of z-coordinate is set to be at the middle of the electron hole separation \cite{Hu2016}.  
Consider the hopping from lattice point $(m,n)$ on the hole layer to $(m+1,n)$ on the electron layer, the phase due to $\mathcal{B}_y$ is 
\begin{eqnarray}
\theta&=&\int_{(m,n,-z_o/2)}^{(m+1,n,z_o/2)}  \vec{A}\cdot dl \notag\\
&=& \int_{(m,n,-z_o/2)}^{(m+1,n,z_o/2)} (\mathcal{B}_yz)dx \notag\\
&=& \int_{(m,n,-z_o/2)}^{(m+1,n,z_o/2,x+a)} (\mathcal{B}_yz) \frac{dx}{dz}{dz}\notag\\
&=& 0
\end{eqnarray}
From the second to the third line, we employed chain rule. $\frac{dx}{dz}$ is a constant and can be pulled out from the integral. Since the vector potential is an odd function along the $z$-direction, the phase is zero.

For completeness, the Hamiltonian in the presence of a tilted magnetic field after the Peierls substitution is

\begin{widetext}
\begin{eqnarray}
H &=& \left(
\begin{array}{cc}
h_+ & 0 \\
0 & h_-
\end{array}
\right), \notag\\
\mbox{, where } 
&h_{\alpha=\pm}&=
\left(
\begin{array}{cc}
M-(B+D)((k_x+\frac{\pi e \mathcal{B}_yz_o}{h}-\frac{e\mathcal{B}_zy}{\hbar})^2+k_y^2 )&  A(\alpha (k_x-\frac{e\mathcal{B}_zy}{\hbar})+ik_y)  \\
 A(\alpha (k_x-\frac{e\mathcal{B}_zy}{\hbar})-ik_y) &-M+ (D-B)((kx-\frac{\pi e \mathcal{B}_yz_o}{h}-\frac{e\mathcal{B}_zy}{\hbar})+k_y^2)
%
\end{array}
\right).
\end{eqnarray}
\end{widetext}

\section{Appendix B: Energy level calculation}
In the presence of an out-of-plane magnetic field $\mathcal{B}_z$, carriers follow cyclotron motion, leading to Landau levels. LLs can be calculated by Peierls substitution. Here we take Landau gauge $\vec{A}=(-\mathcal{B}_zy,0,0)$ which preserves the translation invariance along $x$-direction. We define ladder operators
	\begin{eqnarray}	
	a^{\dagger}&=&\frac{l_B}{\sqrt{2}}(-ik_y-(k_x-e\mathcal{B}_zy))\\
	a&=&\frac{l_B}{\sqrt{2}}(ik_y-(k_x-e\mathcal{B}_zy))
	\label{ladder}
	\end{eqnarray}
 which obey the commutation relation $[a,a^{\dagger}]=1$. The Hamiltonian $H_o$ can be rewritten in terms of the ladder operators as 
 	\begin{eqnarray}
	H_o(a^{\dagger},a)&=& 
	\left(
	\begin{array}{cc}
 	h_+(a^{\dagger},a) &    0 \\
 	0 &     h_-(a^{\dagger},a) 
	\end{array}
	\right),
	\end{eqnarray}	
where 
	\begin{eqnarray}
	h_{\pm}(a^{\dagger},a)&=&\frac{-2D}{l_B^2}(a^{\dagger}a+\frac{1}{2})I+(M-\frac{2B}{l_B^2})(a^{\dagger}a+\frac{1}{2})\sigma_z\notag\\
	&\mp&\frac{\sqrt{2}A}{l_B}(a^{\dagger}\sigma_++a\sigma_-)
	\end{eqnarray}
with $\sigma_{\pm}=(\sigma_x\pm i\sigma_y)/2$.

      As the magnetic field is tilted parallel to the plane, the magnetic field has a y-component. We use the gauge $\vec{A}=(-\mathcal{B}_zy+\mathcal{B}_yz,0,0)$. In terms of the ladder operators as defined in Eq. (\ref{ladder}), the full Hamiltonian becomes $H_o(a^{\dagger},a)+H'(a^{\dagger},a)$, where
      \begin{eqnarray}
		H'(a^{\dagger},a)&=&
		\left(
		\begin{array}{cc}
  		h'(a^{\dagger},a)	& 0    \\
  			0 &   h'(a^{\dagger},a)   
		\end{array}
		\right)\notag\\
		h'(a^{\dagger},a)&=&\frac{e\mathcal{B}_y z_o}{\sqrt{2}l_B}(B I+D\sigma_z)(a^{\dagger}+a) \notag\\
			&-&\left( \frac{e\mathcal{B}_y z_o}{2\hbar} \right)^2(B\sigma_z+D I).
			\label{Hp}
	\end{eqnarray}
	
To calculate its influence, we project $H'(a^{\dagger},a)$  onto the LLs of the $H_o(a^{\dagger},a)$. The first term in Eq. (\ref{Hp}) mixes the adjacent LLs, while the second term affects within each LL. After the projection to the desired number of LLs, we direct diagonalize the full Hamiltonian to obtain the eigenenergies.

\bibliography{TPT_revision_arxiv.bbl}

\end{document}